\documentstyle[twoside,fleqn,espcrc2]{article}
\newcommand{\rf}[1]{(\ref{#1})}
\newcommand{\beq}{\begin{equation}}
\newcommand{\eeq}{\end{equation}}
\newcommand{\bea}{\begin{eqnarray}}
\newcommand{\eea}{\end{eqnarray}}

\renewcommand{\a}{\alpha}

\newcommand{\om}{\omega}

\newcommand{\ra}{\right\rangle}
\newcommand{\la}{\left\langle}

\newcommand{\AmS}{{\protect\the\textfont2
  A\kern-.1667em\lower.5ex\hbox{M}\kern-.125emS}}

\title{Quantum gravity with linear action.\\
               Intrinsic rigidity of spacetime.}

\author{G.K.Savvidy\address{National Research Center "Demokritos",\\ 
        Ag. Paraskevi, GR-15310 Athens, Greece }
        \thanks{Dedicated to Professor Tullio Regge at the occasion of
                his 65th birthday.}}

\begin{document}

\begin{abstract}
An earlier proposed theory with linear-gonihedhic 
action $A(M_{4})$ for quantum gravity,
which requires the existence of a fundamental constant of dimension
one is reviewed. One can consider this theory as a "square root" of classical
gravity. We demonstrate also, that the partition function
for the discretized version of the Einstein-Hilbert action 
found by Regge in 1961 can be represented as a 
$superposition$ of random surfaces with Euler character as an action
and in the case of linear gravity as a superposition of three-dimensional
manifolds with an action which is proportional to the total solid 
angle deficit of these manifolds. This representation allows to construct
the transfer matrix which describes the propagation of space manifold.
We discuss the so called gonihedric principle 
which allows to defind a discrete version of high derivative terms in 
quantum gravity and to introduce generalized deficit angles which 
suppress non-flat vertices, thus introducing intrinsic rigidity of 
spacetime. This note is based on a talk delivered at the II meeting
on constrained dynamics and quantum gravity at Santa Margherita Ligure.

\end{abstract}
 
\maketitle

\section{Introduction}

Unification of gravity with other fundamental forces 
within the superstring theory essentially increased the interest 
to quantum gravity and to physics at Planck scale. In particular,
string theory predicts modification of the gravitational action 
at Planck scale with additional high derivative terms 
and possible terms generated by D-braines. In 
principle one can ask different qustions concerning physics 
at Planck scale refering to this effective action and 
in particular one can try to formulate direct regularization of 
the Euclidian path integral over simplicial manifolds using 
this effective action.

One of the basic elements in this definition of path integral
over simplicial manifolds is the discrete version of
the Einstein-Hilbert action which has been found long time ago by 
Regge \cite{regge}. This action allows to define properly summation 
over simplicial manifolds
\cite{regge,weingarten,adf,david,kkm,kpz,ninomiya,hamber,menotti,av,aj,many}, 
and to
consider nontrivial scaling behaviour of the theory. 

The aim of this article is to review and to extend 
this approach to different modifications of gravity 
which follow from string theory and also to develop an 
alternative approach to quantum gravity 
which is based on new principles 
\cite{savvidy,savvidy1,savvidy2}. These geometrical principles
allow one to extend the notion of 
Feynman integral over paths to an integral over manifolds in the 
sense that the manifold
which degenerates into a single world line
has an amplitude which is proportional to its length. 
This approach to quantum gravity 
was motivated by the fact
that time 
evolution of the space $M^{\tau}$ 
is described by an amplitude which should be proportional
to the linear size of the universe $M$ \cite{savvidy,savvidy2}.
This principle of {\it linearity} together with the 
{\it continuity} principle
in the space of triangulations allows
to define the linear action of the theory $A(M)$. For 
the three-dimensional gravity this linear 
action $A(M_3)$ coinsides with
the Regge action and in four dimensions it is essentially 
different because the Regge action $S(M_{4})$ has dimension 
two and 
is proportional to the area of the universe $M_{4}$ 
while the linear action $A(M_{4})$ is proportional to 
the linear size of the universe and can be considered as a
"square root" of classical gravity. 
The linear character of the theory requires the 
existence of a new fundamental coupling constant of dimension one.
In this theory the convergence of the 
partition function is better and  the linear character
of the theory allows to integrate over the lenghts of the links and 
to sum over different trinagulations at the same time \cite{savvidy2}.
One can expect that quantum fluctuations 
will generate an effective action at 
large distances, which coinsides with 
the classical gravity.

The first part of this review presents the above approach
to quantum gravity, which is based on the idea that quantum mechanical 
amplitudes should be proportional to the "linear size" of the 
fluctuations. It is natural to call these amplitudes "linear" or 
"gonihedric" because their defination contains the sum of 
products of the characteristic lengths and angles of the fluctuations 
\cite{savvidy,savvidy2}.
In the second part we demonstrate that the simplicial quantum 
gravity which is based on the original Regge action can be 
represented as a superposition of topological amplitudes proportional 
to the Euler character. A similar result holds in quantum gravity
based on gonihedric action $A(M_{4})$ where the superposition is over 
total deficit angles. This result allows one to construct the 
transfer matrix
in gravity which describes the propagation of space manifold 
\cite{savvidy2}.
In the third part we review the systematic construction of the 
high derivative terms on a simplicial manifolds and discuss 
the possible consequences on spacetime rigidity. The rigidity 
can be understud as a property
of system when the dynamical dimension in the form of the Hausdorff
dimension $d_{H}$ coincide with the kinematical dimension $d$ 
of the system $d_{H}= d$ \cite{ass} .

\section{Quantum gravity with linear-gonihedric action [19]}

Quantum gravity with linear-gonihedric action $A(M_d)$
can be derived from natural 
physical requirements such as : 

{{\bf $\alpha )$}} coincidence of the transition amplitude 
with the usual Feynman path amplitude for a manifold 
$M_{d}$ degenerated to a single world line and 

{{\bf $\beta )$}} the continuity principle for the transition 
amplitudes.
\vspace{.5cm}

{{\bf d=3}}. To construct the three-dimensional quantum gravity with 
the linear action $A(M_3)$ we shall apply the 
principles $\alpha )$ and $\beta )$.
In accordance with $\alpha )$ the quantum mechanical amplitude should be 
proportional to the {\it linear size of the manifold} and thus  it must be 
proportional to the linear combination of the lengths of all edges of 
simplicial manifold $M_3$

\begin{equation}
A(M_3) = \sum_{<i,j>} \lambda_{ij}\cdot \Theta_{ij}
\end{equation}
where $\lambda_{ij}$ is the length of the edge between 
two vertices $<i>$ and $<j>$, summation is over all
edges $<i,j>$  and
$\Theta_{ij}$ is unknown factor, which can be defined by use of
the continuity principle $\beta )$. Indeed, if we impose a new
vertex $<m>$ inside a given $flat$ tetrahedron $<ijkl>$, then for that 
new manifold we will get additional terms
$\lambda_{im}\Theta_{im}+\lambda_{jm}\Theta_{jm}+
\lambda_{km}\Theta_{km}+
\lambda_{lm}\Theta_{lm}$ to the action and we will get an extra
terms imposing new vertices, despite the fact that the manifold 
does not actually change. To exclude such type of contributions we 
should choose gonimetric factor $\Theta_{ij}$ so that it will vanish
in $flat$ cases. This can be done by the use of the dihedral angles, 
therefore 
\begin{equation}
A(M_3) = \sum_{<i,j>} \lambda_{ij}\cdot
\Theta(~\sum \beta_{ij}) , \label{Hilquan}
\end{equation}
where $\Theta(2\pi) = 0$ and summation is over all 
dihedral angles $\beta_{ij}$  between 
triangular faces of tetrahedra which have the common edge $<i,j>$.
These angles appear in the normal
section of the edge $<i,j>$. This definition guarantees that  flat 
edges, $\sum \beta_{ij}=2\pi$, do not change the action.

In analogy with the gonihedric string \cite{savvidy} 
one can also define more specific theories with the property that 

\begin{equation}
\Theta(4\pi-\sum \beta_{ij}) = 
\Theta(~\sum \beta_{ij}),~~~~~\Theta(~\sum \beta_{ij}) 
\geq  0 ,
\end{equation}
thus we shall allow almost arbitrary functions of 
the deficit angle $\omega^{(2)}_{ij}=2\pi - \sum \beta_{ij}$.
Generally speaking, a suitable selection of the gonimetric
factor $\Theta$ can be done 
if we require a convenient scaling behavior of the 
theory \cite{savvidy}. 
We will use the following parametrization of the $\Theta(\omega)$ 

\begin{equation} 
\Theta(\omega)= ( 2\pi - \omega )^{\varsigma},
\end{equation}
which for the case $\varsigma = 1$ coincides with the 
Regge action \cite{regge}

\begin{equation}
A(M_3) = \sum_{<i,j>} \lambda_{ij}\cdot \omega^{(2)}_{ij}
\label{three}
\end{equation}
and is the discrete version of the following 
continuous Hilbert-Einstein
action

\begin{equation}
A(M_{3})= \int_{M_3}R~ dv_3 . \label{Hil}
\end{equation}

{{\bf d=4}}. Let us consider now four-dimensional manifold 
$M_{4}$. Using the same principles of linearity $\alpha)$ 
and continuity $\beta)$
we can define the linear action $A(M_{4})$
for the four-dimensional quantum gravity as 

\begin{equation}
A(M_{4}) =\sum_{<i,j>}\lambda_{ij}\cdot \sum (2\pi - 
\sum \beta_{ijk}) \label{lingrav}
\end{equation}
where the first summation is extended over 
all edges of $M_{4}$ and the 
second summation is over two-dimensional normal sections 
of a given edge $<i,j>$ and all triangles $<ijk>$ which have 
a common edge $<ij>$,  $\beta_{ijk}$
are the angles on the cone which appear in the normal section of 
the edge $<ij>$ and triangle $<ijk>$.
Geometrically the last 
factor is equal to the total area of the polyhedron on $S^{3}$ 
which corresponds to the spherical image of the edge $<i,j>$, 
that is to total deficit angle on all triangles 
sharing a given edge $<i,j>$. In short notation the action is
\begin{equation}
A(M_{4}) =\sum_{<i,j>}\lambda_{ij}\cdot 
\sum^{triangles~<ijk>~with}_{common~edge~<ij>}~\omega^{(2)}_{ijk}
\label{lamb}
\end{equation}
where $\omega^{(2)}_{ijk}$ is the deficit angle accociated with
triangle $<ijk>$. 

Let as combine all terms in (\ref{lamb}) belonging to a given triangle
$<ijk>$, then the sum $\lambda_{ij}+\lambda_{jk}+\lambda_{ki}=
\lambda_{ijk}$ is the perimeter of the triangle $<ijk>$, 
thus
\begin{equation}
A(M_{4}) = \sum_{<ijk>}\lambda_{ijk} \cdot \omega^{(2)}_{ijk}.
\label{lamb1}
\end{equation}
This is an equivalent 
form of the action (\ref{lamb}), therefore one 
can multiply
the length of the edge by the 
total area of the polyhedron on $S^{3}$ which corresponds to 
the spherical image  of that edge or 
the perimeter of the given triangle by the area of the 
image on $S^{3}$
which corresponds to that triangle (\ref{lamb1}).
The linear character of the theory requires the 
existence of a new fundamental coupling constant of dimension one.
This linear theory is again $intrinsic$ 
and has better chances to describe quantum gravity because 
the entropy factor is finite in this case.

The difference with the classical area action 
is that in (\ref{lamb1}) the {\it perimeter}  
of a triangle is  multiplied by the
deficit angle associated with the triangle 
and in the Regge action  one should multiply 
the {\it area} of the  
triangle by the corresponding deficit angle

\begin{equation}
S(M_{4})= \sum_{<ijk>} \sigma_{ijk} \cdot \omega^{(2)}_{ijk},
\label{area}
\end{equation}
the summation is over all 
triangles in both cases. 

The third equivalent form of the linear action (\ref{lamb}) 
(\ref{lamb1}) can be found by using Gauss-Bonne theorem.
Indeed at the vertex $<i>$ of a three-dimensional simplex we 
have 
\cite{aw,herglotz,allendoerfer,santalo,santalo1,schneider}
\begin{equation}
\frac{1}{8\pi}\sum \omega^{(2)}_{ij} + 
\frac{1}{8\pi}\sum \Omega^{(3)}_{ijkl} = \frac{1}{2}
\end{equation}
where $\sum \omega^{(2)}_{ij}$ is the total deficit angle 
associated with all edges $<ij>$ with a common
vertex $<i>$ and 
$\sum \Omega^{(3)}_{ijkl}$ is the total solid angle of all 
tetrahedra $<ijkl>$ adjacent to a vertex $<i>$. Let us consider 
the normal section of the edge $<ij>$ of the four-dimensional 
manifold $M_{4}$ and then apply the above theorem to a three-
dimensional vertex which appears in the normal section

\begin{eqnarray}
\sum^{triangles~<ijk>~with}_{common~edge~<ij>} \omega^{(2)}_{ijk}
= 4\pi -
\sum^{tetrahedra~<ijkl>~with}_{common~edge~<ij>} \Omega^{(3)}_{ijkl}
\end{eqnarray}
where $4\pi -\sum \Omega^{(3)}_{ijkl}\equiv \Omega^{(3)}_{ij}$ 
is the $solid$ angle deficit associated with edge
$<ij>$. Thus the gonihedric action (\ref{lamb}) can be 
rewritten in the form

\begin{equation}
A(M_4) = \sum_{<ij>} \lambda_{ij} \cdot \Omega^{(3)}_{ij}.
\label{solid}
\end{equation}
This is a new result and we will use this expression in 
the next section to 
represent linear gravity
in the form of superposition of topological amplitudes.

In his basic paper on the subject Regge (1961) notes that one
can derive Einstein's equations in empty space by the 
variation of the action over lengths of the edges 
$\delta \lambda_{ij}$ as if the deficit angles $\omega^{(2)}_{ijk}$
were constant $\delta~\omega^{(2)}_{ijk}=0 $ . 
The same result holds in our case, therefore 
we can find equations for the gravity with gonihedric action 
(\ref{lamb1})
\begin{equation}
\frac{\delta~A(M_{4})}{\delta~\lambda_{ij}} = 
\sum^{triangles~with}_{common~edge~<ij>}~\omega^{(2)}_{ijk}=0
\end{equation}
It is usefull to compare this equations with the Regge 
equations
\begin{equation}
\frac{\delta~A(M_{4})}{\delta~\lambda_{ij}} =
\sum^{triangles}_{common~edge~<ij>}\omega^{(2)}_{ijk}
ctg~\gamma_{ijk}=0
\end{equation}
where $\gamma_{ijk}$ is the angle which is opposite to the edge
$<ij>$ in the triangle $<ijk>$. As we have seen the equations
are slightly simpler than in the classical general relativity 
but still nontrivial in the sence that the spacetime 
is not simply flat as is the case of three-dimensional 
gravity where the equations have the form
\begin{equation}
\frac{\delta~A(M_{3})}{\delta~\lambda_{ij}} = \omega^{(2)}_{ijk}=0
\end{equation}
that is three dimensional Einstein spacetime is flat.

The partition function of the simplicial quantum gravity 
with linear action (\ref{lamb1})
can be represented in the form

\begin{equation}
Z_{gravity}(\beta) \approx  \sum_{\{ M_4 \}} 
\exp\{- \beta A(M_4)\} \label{Z}
\end{equation}
and one can expect that quantum fluctuations will generate effective 
action which is proportional to Einstein-Hilbert action

\begin{equation}
Z_{gravity}(\beta) = exp\{ -\beta A(M_4) - 
\frac{1}{G(\beta)}S(M_4)-\cdot \}
\end{equation}

\section{Quantum gravity as a superposition of 
topological amplitudes [19,33]} 

Our aim now is to represent quantum gravity 
with the action (\ref{three}) and (\ref{area}) as a 
$superposition$ of less complicated geometrical theory 
of random surfaces with Euler character as an action

\begin{equation}
\chi(M_2) = \sum_{<ver>} \omega^{(2)}_{ver}
\label{char}
\end{equation}
where $\omega^{(2)}_{ver}$ is the deficit angle associated  with
the vertex and summation is over all verteces of the surface $M_{2}$.
Let us consider the intersection of the simplicial manifold 
$M_3$  which is embedded into Euclidean space of arbitrary large 
dimension $d$
by the $d-1$ dimensional plane $E$. The intersection is 
the two-dimensional surface $M^{E}_{2}$

\begin{equation}
M^{E}_{2} = M_{3} \cap E ,
\end{equation}
because $dim M^{E}_{2}= dim M_{3} +dim E -dim R^{d}=3 +(d-1) -d =2$.

We will see now that the linear action $A(M_3)$ is 
the sum of 
Euler characters of all surfaces which appear in the intersection
$\{ M^{E}_{2}\}$

\begin{equation}
A(M_3) = \sum_{\{ E\}}~\chi(M^{E}_{2}~) , \label{super}
\end{equation}
where 

\begin{equation}
\chi(M^{E}_{2}) = \sum_{<i,j>} \omega^{(2)E}_{ij}.
\label{chare}
\end{equation}
and $\omega^{(2)E}_{ij}=2\pi-\sum\beta^{E}_{ij}$,~ $\beta^{E}_{ij}$ 
are the 
angles in the intersection of the plane $E$ with the edge $<i,j>$
and $\omega^{E}_{ij}= 0$ for the edges of $M_3$ which are not 
intersected by the given plane $E$. 

The formula (\ref{super}) follows from the fact that 
the average of the angle $\beta^{E}_{ij}$ over 
all intersecting planes $\{ E \}$ is equal to dihedral angle 
$\beta_{ij}$
and that the number of planes which intersect the given edge $<i,j>$
is proportional to its length $\lambda_{ij}$ 
\cite{poincare,blaschke,chern,sch,savvidy2}

\begin{equation}
<\beta^{E}_{ij}> \equiv \int_{E^{d-1}} \beta^{E}_{ij}~ dE 
= \beta_{ij} 
\cdot \lambda_{ij} .\label{avar}
\end{equation} 
Therefore we 
have the following representation of the partition function 
of the three-dimensional quantum gravity

\begin{equation}
Z_{gravity}(\beta)  
= \sum_{\{M_3\}}~ \prod_{\{ E \}}~ exp\{-\beta~\chi(M^{E}_{2})~\}.
\label{statsuper}
\end{equation}
In deriving this result we supposed that  the universe had been 
embedded into a flat Euclidean space of {\it arbitrary} large dimension 
$d$ , which seems not to be a very strong restriction, but still 
necessary element in this construction. 

In four dimensions the action (\ref{area})  is proportional 
to the area 
$S(M_{4})$ of the four-dimensional universe $M_{4}$ 
and we shall again assume that it is embedded 
into Euclidean space of an
arbitrary large  dimension $d$. 
Then the intersection (the tomography)
of the universe $M_{4}\subset R^{d}$ by the 
(d-2)-dimensional plane $E$ is the two dimensional 
surface $(dim M^{E}_{2} =
dim M_{4} + dim E - dim R^{d}
=4 + (d-2) -d =2 )$. The same 
arguments as before allow to show that

\begin{equation}
S(M_{4}) = \sum_{\{ E\}}~\chi(M^{E}_{2}~) \label{areagrav}, 
\end{equation}
where the summation is extended over all $d-2$ dimensional planes 
$\{E\}$ and we have used the fact that

\begin{equation}
<\beta^{E}_{ijk}> \equiv \int_{E^{d-2}} \beta^{E}_{ijk}~ dE 
= \beta_{ijk} 
\cdot \sigma_{ijk} .
\end{equation}
Thus  the action $S(M_{4})$ 
(\ref{area}) is 
equal to the sum of Euler characters $\chi (M^{E}_{2})$ over 
all intersecting planes $\{E\}$. 
With this result we have the same representation 
(\ref{statsuper}) for the 
partition function of the four-dimensional quantum gravity.

This result has a general nature and can be extended to high 
dimensions as well. Indeed the Regge action for $k$-dimensional
universe $M_{k}$ has the form

\begin{equation}
V(M_{k}) =\sum_{<i_1...i_{k-1}>}~v_{i_1...i_{k-1}}~
\omega^{(2)}_{i_1...i_{k-1}} \label{hilbert}
\end{equation}
where $v_{i_1...i_{k-1}}$ is the volume 
element of the $k-2$-dimensional 
subsimplex of 
the $k$-dimensional simplex $<i_1...i_{k}>$ and 
$\omega^{(2)}_{i_1...i_{k-1}}$ is the deficit angle 
associated with the face $<i_1...i_{k-1}>$.

As in low dimensional cases we will consider $M_{k}$ 
embedeed into Euclidean space of arbitrary large dimension
$d$. Intersection of $M_{k}$ by the $d-k+2$-dimensional 
plane is a two dimensional surface $M^{E}_{2}$
$(dim M^{E}_{2} = dim M_{k} + dim E_{d-k+2} - dim R^{d}
=k + (d-k+2) -d =2 )$ and the Euler character of that 
surface is equal to 

\begin{equation}
\chi(M^{E}_{2}) = \sum_{<i_1...i_{k-1}>} 
\omega^{(2)E}_{i_1...i_{k-1}} 
\label{}
\end{equation}
where $\omega^{(2)E}_{i_1...i_{k-1}}=
2\pi-\sum\beta^{E}_{i_1...i_{k-1}}$,~~ $\beta^{E}_{i_1...i_{k-1}}$ 
are the 
angles in the intersection of the plane $E$ with the 
subsimplex $<i_1...i_{k-1}>$ and 

\begin{eqnarray}
<\beta^{E}_{i_1...i_{k-1}}> \equiv \int_{E_{d-k+2}} 
\beta^{E}_{i_1...i_{k-1}}~ dE \\
= \beta_{i_1...i_{k-1}} 
\cdot v_{i_1...i_{k-1}} 
\end{eqnarray}
therefore

\begin{equation}
V(M_k) = \sum_{\{ E\}}~\chi(M^{E}_{2}~) . \label{}
\end{equation}
The partition function of the $k$-dimensional gravity with
action (\ref{three}),(\ref{area}) and 
(\ref{hilbert}) can be represented now in the form

\begin{equation}
Z^{k}_{gravity}(\beta)  
= \sum_{\{M_k\}}~ \prod_{\{ E_{d-k+2} \}}~ exp\{-\beta~\chi(M^{E}_{2})~\}.
\label{statsuperk}
\end{equation}
The similar result holds for gravity with linear action 
(\ref{lamb}),(\ref{lamb1}) and (\ref{solid}).
The intersection of $M_{4}$ by the $d-1$ dimensional plane $E$
is a three-dimensional manifold $M^{E}_{3} = M_{4} \cap E_{d-1}$ ,
because $dim M^{E}_{3}=4 +(d-1) -d =3$. Introducing the integral 
invariant which is equal to total solid angle deficit of the manifold 
$M^{E}_{3}$ 

\begin{equation}
\Omega(M^{E}_3) = \sum_{<ij>}~\Omega^{(3)E}_{ij}  \label{}
\end{equation}
we get 

\begin{equation}
A(M_4) = \sum_{\{E_{d-1} \}} \Omega(M^{E}_{3})
\label{}
\end{equation}
which coinsides with (\ref{solid}) if we use the formula

\begin{equation}
<\Omega^{(3)E}_{ij}> \equiv \int_{E_{d-1}} 
\Omega^{(3)E}_{ij}~ dE \\
= \Omega^{(3)}_{ij} 
\cdot \lambda_{ij} 
\end{equation}
Thus in the heart of the general relativity is the universal topological
theory with Euler character as an action and the new invariant in the 
case of linear gravity.

\section{Transfer matrix [18,19]}

In (\ref{statsuper}) and (\ref{statsuperk})
the summation is not over independent two-dimensional surfaces 
$\{M^{E}_{2}\}$, because these surfaces "remember" a parent 
manifold $M_{k}$, from which they appear in the prosses of slicing.
The problem is to pass  from summation over 
two-dimentional surfaces $M^{E}_{2}$ to the summation 
over independent surfaces $\{M^{\tau}_{2}\}$ the union of 
which can "reproduce"
any random manifold $\{M_{k}\}$. 
These problem (constraints) can be 
resolved geometrically if we consider the system on the lattice.
When the continuous Euclidean space is replaced by the Euclidean
lattice, where the surfaces and the manifolds are associated with
the collection of the plaquettes and cubes, then
the product over all intersecting planes 
$\{E\}$ can be evaluated 
to a product over planes $\{E^{\tau}\}$ which are perpendicular 
to a given time direction $\tau$ 

\begin{equation}
Z_{gravity}(\beta) = 
\sum_{\{..M^{\tau}_{2},M^{\tau+1}_{2}..\}}~ \prod_{\tau}~K(M^{\tau}_{2},
M^{\tau+1}_{2})
\label{DiGrav},
\end{equation}
where

\begin{eqnarray}
K(M^{\tau}_{2},M^{\tau +1}_{2}) = 
exp -\beta \{ \frac{1}{2} \Theta(M^{\tau}_{2}) + A(M^{\tau}_{2}) 
+\\
\frac{1}{2} \Theta(M^{\tau+1}_{2}) + A(M^{\tau +1}_{2}) 
-2 A(M^{\tau}_{2} \cap M^{\tau +1}_{2})~ \} \label{FeyGrav}
\end{eqnarray}
and the $independent$ summation is extended 
over all surfaces $\{.. M^{\tau}_{2},M^{\tau+1}_{2}.. \}$ on 
different time slices. 
We have the propagation of the "space" surface 
$M^{\tau}_{2}$ in the time direction $\tau$ 
with an amplitude which is proportional to the sum of the 
generalized Euler 
character $\Theta(M^{\tau}_{2})$ and 
of the linear size of the space surface 
$A(M^{\tau}_{2})$ which is defined as 

\begin{equation}
A(M_2) = \sum_{<i,j>}  \lambda_{i,j}
\cdot \vert \pi -\alpha_{ij}\vert 
\end{equation}
where $\alpha_{i,j}$~ is the dihedral angle between two neighbor
faces of $M_2$ having a common edge $<i,j>$ of the length 
$\lambda_{i,j}$. The interaction is proportional to the 
length of the right angle edges of the 
overlapping  surface $A(M^{\tau}_{2} \cap M^{\tau+1}_{2})$.
This formula is valid when 
$\Theta(\omega)=\vert 2\pi - \omega \vert $,
the embedding space has dimension four 
and self-intersection coupling constant $k$ is equal to 
infinity \cite{savvidy2}. 

\section{Gonihedric principle and intrinsic rigidity 
of spacetime [20]}
In this section we will describe the  regular way of constructing
the integral invariants on a simplicial manifold which naturally 
reproduces the Regge result and allows to construct large 
class of {\it new integral invariants} which can serve in our attempt to 
formulate well defined path integral in quantum gravity. 
 
The method is based on the Steiner idea of parallel manifold 
\cite{steiner,weyl,ambarzum}. The hypervolume of the parallel manifold 
$M^{\rho}_{n-1}$ can be expanded into a polynomial of a distance 
$\rho$ from the original manifold $M_{n-1}$. Every term of this expansion 
represents an integral invariant $\mu_{k}(M_{n-1})$~~~~$k=0,1,...,n-1$
constructed on the $n-1$-dimensional manifold $M_{n-1}$ through the 
curvature tensor \cite{weyl}. 

Let us consider for that a compact orientable hypersurface $M_{n-1}$
embedded in an Euclidean space $E^{n}$ and define 
a parallel manifold $M^{\rho}_{n-1}$ as a set of all 
points at a distance $\rho$ from $M_{n-1}$. Then for 
$\rho$ sufficiently small the hypervolume $\mu_{0}(M^{\rho}_{n-1})$
of the parallel manifold is equal to 

\begin{equation}
\mu_{0}(M^{\rho}_{n-1}) = \int  (R_{1}+\rho)...
(R_{n-1}+\rho) 
d \omega_{n-1} \label{expa},
\end{equation}
simply because $n-1$ principal curvatures for the parallel 
manifold $M^{\rho}_{n-1}$ are equal to $R_{i}+\rho$,~~~$i=
1,2,..,n-1$. Expanding the product of the integrand 
one can get 
\begin{eqnarray}
\mu_{0}(M^{\rho}_{n-1}) = \mu_{0}(M_{n-1}) +\rho\cdot \mu_{1}(M_{n-1}) 
\nonumber\\ + \rho^{2}\cdot \mu_{2}(M_{n-1})+...
+ \rho^{n-1}\cdot \mu_{n-1}(M_{n-1})     
         \label{invari},
\end{eqnarray}
thus generating the whole sequence of integral invariants
$\mu_{k}(M_{n-1})=\int\{R_{i_{1}}...R_{i_{n-k-1}} \}d\omega_{n-1}$.

The main idea of the construction of the discrete invariants is based 
on the fact that this expansion can be evaluated not only 
for smooth manifolds but for the simplicial manifolds as well.
This allows to find out discrete versions of the above classical 
$\mu_{k}(M_{n-1})$ invariants which includes also the Hilbert-Einstein 
action $\mu_{2}(M_{n-1})$.

To demonstrate the method let us consider Steiner expansion
for the hyper-volume $\Omega$ of a smooth manifold $M_{4}$  
\begin{eqnarray}
\Omega(M^{\rho}_{4}) &=& \Omega(M_{4}) +\rho \cdot V(M_{4})
+ \rho^{2} \cdot S(M_{4}) \nonumber\\ &+&\rho^{3} \cdot A(M_{4}) +
\rho^{4}\cdot \chi(M_{4})
\label{}
\end{eqnarray}
and compare with the same expression for a piecewise linear manifold
$M_{4}$, then one can get the discrete versions of the above invariants
\cite{ass}
\begin{eqnarray}
\Omega(M_{4}) &=& \sum_{\la ijklm\ra } v_{ijklm} \cdot 1,\label{hype}\\
V(M_{4})&=& \sum_{\la ijkl\ra } v_{ijkl} \cdot ( \pi - \alpha_{ijkl}),
\label{volu}\\
S(M_{4})&=& \sum_{\la ijk\ra } \sigma_{ijk} \cdot ( 2\pi - \sum \beta_{ijk}),
\label{areah}\\
A(M_{4})&=& \sum_{\la ij\ra } \lambda_{ij} \cdot
\omega^{(3)}_{ij}(\alpha,\beta),
\label{lineh}\\
\chi(M_{4})&=& \sum_{\la i\ra }1\cdot \omega^{(4)}_{i}(\beta), 
\label{disch}
\end{eqnarray}
where we introduce internal angles $\beta_{i}, \beta_{ij},
\beta_{ijk}$ between one, two and three simplexes and $\alpha_{ijkl}$
for the external angle between two four-dimensional simplexes,
as well as the notation $v_{ijklm}$ for the four-volume of the
four-simplex $\la ijklm\ra$ and the notation $\om^{(4)}_i$ for the
four-volume in $S^4$ of the spherical image of the vertex $\la i \ra$.

Note that $\omega^4_i$ is independent of the external angle $\a$.
This is in agreement with the general pattern 
which states that the coefficients to even powers
of $\rho$ are intrinsic integral invariants, while the
coefficients to the odd powers of $\rho$ contain reference to the
extrinsic geometry.
The whole four-volume (\ref{disch}) obtained by summation over
all vertices of the piecewise linear manifold $M_4$
is proportional to the Euler-Poincare
character in the same way as in the two-dimensional case
(\ref{char}).

The important lesson which we learn from Steiner expansion 
for smooth and non-smooth triangulated manifolds 
(\ref{hype})-(\ref{disch}) is that
in all these cases the integral invariants $\mu_{k}(M_{n-1})$ 
are the product of the $volume$ of the faces of $M_{n-1}$ 
and the $volume$ of the corresponding normal images
of these faces 
\begin{equation}
\mu = \sum_{\{faces\}} (vol~of~face ) \cdot 
(vol~of~image) \label{fact}
\end{equation}
here the "face" means vertex, edge, triangle, tetrahedron .. 
and high dimensional sub-simplexes of $M_{n-1}$.

Using the gonihedric property (\ref{fact}) of the
integral invariants 
one can construct new invariants on 
a manifold $M_{n-1}$ \cite{ass}. To maintain locality one should always 
multiply the $volume$ of the face by the different 
$invariant~measures$ on the corresponding image
\begin{equation}
\sum_{\{faces\}} (vol~of~face ) \cdot 
(measure~on~image). \label{goni}
\end{equation}
The only restriction is
that the topological measures on the spherical images should be 
$excluded$, 
because a given vertex can be deformed into a flat one, in which 
case the spherical image shrinks to zero leaving nonzero topological
invariant. Therefore one should always exclude topological measures 
on a spherical images when exploring new invariants \cite{ass}.

To analyze  the possible new integral invariants
in four and higher dimensions, we
need to introduce a universal notation for {\it solid angles} on
the faces of a simplex and on the corresponding spherical images. We shall
use $\Omega^{(k)}$ for solid angle at a vertex of a
$k$-dimensional simplex  and
$\omega^{(k)}$ for the solid angles on spherical images.
All these solid angles are functions of the previously introduced angles
denoted by $\alpha$'s and  $\beta$'s
with various indices.

The geometric measures on the spherical image of
the vertex $\la i\ra $ are: hyper-volume, volume, area and length
\cite{ass}
\begin{eqnarray}
\omega^{(4)}_{i},~~~~~~~\sum\omega^{(3)}_{ij},\nonumber\\
\sum\Omega^{(2)}_{ijk}~\omega^{(2)}_{ijk},~~~~~~~\sum
\Omega^{(3)}_{ijkl}~\omega^{(1)}_{ijkl}.
\end{eqnarray}
Using these measures one can construct new invariant of the ``area''
type (like  $S(M_4)$), namely
\begin{equation}
\mu_{S}(M_{4})=\sum_{\la ijk\ra } \sigma_{ijk}
\cdot (~ \sum ~\omega^{(1)}_{ijkl}~) \label{fams}.
\end{equation}
There are two new invariants  of ``length'' type (i.e. like $A(M_4)$):
\begin{equation}
\mu_{A}(M_{4})=\sum_{\la ij\ra } \lambda_{ij} \cdot \left\{ \begin{array}{c}
\sum ~\omega^{(2)}_{ijk}\\
\sum~\Omega^{(3)}_{ijkl}~\omega^{(1)}_{ijkl}, \label{fama}
\end{array} \right\}
\end{equation}
and finally three new invariants which are dimensionless,
(like $\chi(M_4)$):
\begin{equation}
\mu_{\chi}(M_{4}) = \sum_{\la i\ra } 1 \cdot \left\{ \begin{array}{c}
\sum ~\omega^{(3)}_{ij}\\
\sum~\Omega^{(2)}_{ijk}~\omega^{(2)}_{ijk}\\
\sum~\Omega^{(3)}_{ijkl}~\omega^{(1)}_{ijkl} \label{famx}
\end{array} \right\}
\end{equation}
We shall perform now the dual transformation
of the above invariants and address the question of their internal
and external properties.

To get the dual form of the above invariants we combine terms
in the parentheses belonging to the same tetrahedron in the first,
third and sixth of the invariants in \rf{fams}--\rf{famx}:
\begin{equation}
\left\{ \begin{array}{c}
\sigma_{ijkl} \\
\lambda_{ijkl}\\
\Omega_{ijkl}
\end{array} \right\}  = \sum_{over~tetrahedron~\la ijkl\ra }
\left\{ \begin{array}{c}
\sigma_{ijk}\\
\lambda_{ij}~\Omega^{(3)}_{ijkl}\\
\Omega^{(3)}_{ijkl}
\end{array} \right\}.
\end{equation}
Hence, we have the following sequence of invariants constructed from
$\omega^{(1)}_{ijkl}$ associated with tetrahedron $\la ijkl\ra$:
\begin{equation}
\mu_{\omega}(M_{4}) = \sum_{\la ijkl\ra } \left\{ \begin{array}{c}
v_{ijkl}\\
\sigma_{ijkl}\\
\lambda_{ijkl}\\
\Omega_{ijkl}
\end{array} \right\}  \cdot \omega^{(1)}_{ijkl}, \label{ome1}
\end{equation}
where $\omega^{(1)}_{ijkl}=\pi - \alpha_{ijkl}$ is the angle between
two neighboring  four-dimensional simplexes having a
common tetrahedron $\la ijkl\ra $
of volume $v_{ijkl}$, area $\sigma_{ijkl}$, length $\lambda_{ijkl}$
and total internal solid angle $\Omega_{ijkl}$.

Combining the second and the fifth term in the parentheses in \rf{fams}
and \rf{famx} belonging to the same triangle
\begin{equation}
\left\{ \begin{array}{c}
\lambda_{ijk}\\
\pi
\end{array} \right\}  = \sum_{over~triangle~\la ijk\ra }
\left\{ \begin{array}{c}
\lambda_{ij}\\
\Omega^{(2)}_{ijk}
\end{array} \right\}
\end{equation}
we get integral invariants constructed from  $\omega^{(2)}_{ijk}$'s
associated to triangles $<ijk>$:
\begin{equation}
\mu_{\omega}(M_{4}) =~~\sum_{\la ijk\ra }~~ \left\{ \begin{array}{c}
\sigma_{ijk}\\
\lambda_{ijk}\\
\pi
\end{array} \right\}~~ \cdot \omega^{(2)}_{ijk}, \label{ome2}
\end{equation}
where $\omega^{(2)}_{ijk}=2\pi - \sum \beta_{ijk}$ is the deficit angle
on the triangle $\la ijk\ra $ which have the area $\sigma_{ijk}$, the
perimeter $\lambda_{ijk}$ and as usually the sum of internal angles
equal to $\pi$. The first invariant in this family
coincides with discrete version of the Hilbert-Einstein action
\cite{regge}, the second coincides with the linear action which we 
already discussed  in the first section  
\cite{savvidy2} and the last one is equal to the total deficit
angle of the whole manifold \cite{ass}.

The fourth invariant is simply equal to the volume of the spherical image
of the edge and associated to the linear term (\ref{lineh}) we have
two integral invariants constructed from the three-volume of the
spherical image $\omega^{(3)}_{ij}$ of the edge $\la ij \ra$:
\begin{equation}
\mu_{\omega}(M_{4}) =~~\sum_{\la ij\ra }~~ \left\{ \begin{array}{c}
\lambda_{ij}\\
1
\end{array}
\right\}~~ \cdot \omega^{(3)}_{ij}. \label{ome3}
\end{equation}
Because the spherical images $\omega^{(k)}$ are 
defined through the embedding of 
the manifold into Euclidean space it is not an easy task
to understand why part of the integral invariants (\ref{ome1}),
(\ref{ome2}),(\ref{ome3}) and (\ref{disch}) does not 
actually depend on the embedding. The reason is that, for the smooth 
manifold the 
integral invariants are built up by the Riemann tensor and thus 
are independent of its embedding in Euclidean space. For the 
discrete version of the classical invariants and for the new ones 
this can be seen by applying the  Gauss-Bonnet theorem to a 
spherical images \cite{ass} 
and expressing the measures on the image $\omega^{(k)}$ 
through the internal angles . 

The Gauss-Bonnet theorem provides us with 
general relations between the volume of
the spherical image $\omega^{(k)}$ of the
vertex and the solid angles $\Omega^{(k)}$ associated to the vertex.
Indeed, the Gauss-Bonnet theorem for the two-dimensional
triangulated surface can be
formulated in the form 
\cite{aw,herglotz,allendoerfer,santalo,santalo1,schneider}
\begin{equation}
\frac{1}{4\pi}\omega^{(2)}_{i} + \sum\frac{1}{4\pi}\Omega^{(2)}_{ijk}=
\frac{1}{2}.
\end{equation}
Hence, summing over all vertices in the triangulation one can see that
\begin{eqnarray}
\chi(M_{2}) \equiv \frac{1}{2\pi}\sum_{\la i\ra}\omega^{(2)}_{i}
\\= \sum_{\la i\ra}\{1-\frac{1}{2\pi}\sum \Omega^{(2)}_{ijk}\}
= N_{0} - N_{2}/2 ,
\end{eqnarray}
where $N_{0}$ is the number of vertices and $N_{2}$ the number of
triangles on the surface $M_{2}$.

This can be generalised to four dimensions:
in every vertex of the simplicial
manifold $M_{4}$ we have 
\cite{aw,herglotz,allendoerfer,santalo,santalo1,schneider,che} 
\begin{equation}
\frac{3}{8\pi^{2}}\omega^{(4)}_{i} + \frac{1}{8\pi^{2}}
\sum \Omega^{(2)}_{ijk}~\omega^{(2)}_{ijk} + \frac{1}{4\pi^{2}}
\sum \Omega^{(4)}_{ijklm}=\frac{1}{2},
\end{equation}
which shows that the hyper-volume $\omega^{(4)}_{i}$ on $S^4$ 
can be expressed
through the intrinsic  quantities.
The Euler-Poincare character is equal to
\begin{eqnarray}
\chi(M_{4}) \equiv \frac{3}{4\pi^{2}}\sum_{\la i\ra}\omega^{(4)}_{i}
= \sum_{\la i\ra} \{ 1 - \frac{1}{2\pi^{2}}\sum \Omega^{(4)}_{ijklm} \}
\nonumber\\
- \frac{1}{2}\cdot \frac{1}{2\pi}\sum_{\la ijk\ra}\omega^{(2)}_{ijk} =
N_{0}-N_{2}/2 ~ +N_{4}
\end{eqnarray}
and we obtain the  relation between the total deficit angle
\begin{equation}
\omega^{(2)}_{tot}\equiv \frac{1}{2\pi}\sum_{\la ijk\ra}\omega^{(2)}_{ijk}
\end{equation}
and the total solid deficit angle

\begin{equation}
\Omega^{(4)}_{tot}= \sum_{\la i\ra}(1-\frac{1}{2\pi^2}
\Omega^{(4)}_i) \equiv \sum_{\la i\ra}
\{ 1 - \frac{1}{2\pi^{2}}\sum \Omega^{(4)}_{ijklm} \}
\end{equation}
in the form  \cite{rw}
\begin{equation}
\chi(M_{4}) = \Omega^{(4)}_{tot} - \frac{1}{2}\omega^{(2)}_{tot}.
\label{disvers}
\end{equation}
The Euler-Poincare character is 
\begin{equation}
\chi(M_4) = {1\over 128 \pi^2}\int_{M_4} 
dv_4 \Bigl(R^{2}_{\mu\nu\lambda\rho} -4 R^{2}_{\mu\nu} + R^{2}\Bigr),
\label{inteuler}
\end{equation}
and comparing with the discretized version \rf{disvers} of $\chi(M_4)$
it is not unnatural to associate higher powers of the
integral of curvature tensors with linear combinations of
$\Omega^{(4)}_{tot}$ and $\omega^{(2)}_{tot}$.

From these considerations it seems that the general
discretized action in simplicial
four-dimensional gravity should be of the form:
\cite{ass}
\begin{eqnarray}
Action = \frac{1}{G}\sum_{\la ijk\ra } \sigma_{ijk} 
\cdot \omega^{(2)}_{ijk} \nonumber\\
+ g_1\sum_{\la ijk\ra}\omega^{(2)}_{ijk}+   
g_2\sum_{\la ijk\ra}\Bigl(\omega^{(2)}_{ijk}\Bigr)^2+ ... \nonumber\\
k_1 \sum_{\la i\ra } \Bigl(1-\frac{1}{2\pi^2}\Omega^{(4)}_i\Bigr) +
k_2 \sum_{\la i\ra} \Bigl(1-\frac{1}{2\pi^2}\Omega^{(4)}_i\Bigr)^2 + ... 
\label{genaction}
\end{eqnarray}
where the terms involving the square of the deficit angles
introduce an intrinsic rigidity into the simplicial manifolds.
This choice of discretized action with higher derivative terms
is related, but not identical to the terms suggested in \cite{hw1}.
Generally spiking the rigidity can be understud 
as a property
of system when the dynamical dimension in the form of the Hausdorff
dimension $d_{H}$ coincide with the kinematical dimension $d$ 
of the system \cite{ass}
\begin{equation}
d_{H}= d .
\end{equation}
From numerical simulations one can conclude that the correct 
disription of the 
four dimensional universe in the infrared limit should include terms 
which introduce additional rigidity into quantum gravity.

\section{Acknowledgement}

I wish to thank the organizers of the conference for inviting me
and for arranging an interesting and stimulating meeting. I also thank
J.Ambj\o rn,  K.Savvidy and R.Schneider for an enjoyble collaboration on 
the work reported here and E.Argyres for his kind support.


\begin{thebibliography}{99}


\bibitem{regge}T.Regge, Nuovo Cimento 19 (1961) 558
\bibitem{sorkin}J.A.Wheeler, "Regge Calculus and Schwarzschild Geometry"
in Relativity, Groups and Topology ed. B. DeWitt and C. DeWitt 
(New York: Gordon and Breach, 1964)  pp 463-501\\
R.Sorkin. Phys.Rev.D12 (1975) 385;\\
M.Lehto, H.B.Nielsen and M.Ninomiya. Nucl.Phys. B272 (1986) 228
\bibitem{weingarten}D.Weingarten, Nucl.Phys. B210 (1982) 229
\bibitem{adf} J. Ambj\o rn, B. Durhuus, J. Fr\"{o}hlich and P. Orland:\\
 Nucl.Phys. { B257} (1985) 433;  { B270} (1986) 457;
 { B275} (1986) 161-184.
\bibitem{david} F. David, Nucl.Phys. B257 (1985) 543.
\bibitem{kkm}V.A. Kazakov, I. Kostov and A.A. Migdal, Phys.Lett. B157
(1985) 295; Nucl.Phys. B275 (1986) 641.
\bibitem{kpz}V. Knizhnik, A. Polyakov and A. Zamolodchikov, Mod.Phys.Lett
             A3 (1988) 819.
\bibitem{ninomiya}A. Jevicki and M. Ninomiya, Phys.Rev.D33 (1986) 1634. 
\bibitem{hamber}H.W. Hamber, Nucl.Phys. B (proc.supll.) 25A (1992) 150.
\bibitem{menotti}P.Menotti and P.P.Peirano, Nucl.Phys.B (Proc.Suppl.) 
47 (1996) 633
\bibitem{av} 
J. Ambj\o rn, B. Durhuus and T. Jonsson. Mod.Phys.Lett.
A6 (1991) 1133.\\
J. Ambj\o rn and S. Varsted,
Phys.Lett. { B266} (1991) 285; 
 Nucl.Phys. { B373} (1992) 557.\\
M.E. Agishtein and A.A. Migdal, Mod. Phys. Lett. A6 (1991) 1863.\\
B.Boulatov and A. Krzywicki, Mod.Phys.Lett A6 (1991) 3005.\\
J. Ambj\o rn, D.V. Boulatov, A. Krzywicki and S. Varsted,
Phys.Lett. { B276} (1992) 432.
N.Sakura, Mod.Phys.Lett. A6 (1991) 2613.\\
N.Godfrey and M.Gross. Phys.Rev. D43 (1991) R1749.\\
S.Catterall,J.Kogut and R.Renken. Phys.Lett. B342 (1995) 53.
\bibitem{witt}E.Witten. Nucl.Phys. B311 (1988) 96 ; B323 (1989) 113
\bibitem{thooft}G 't Hooft. Class.Quantum Grav. 13 (1996) 1023
\bibitem{aj}J. Ambj\o rn and J.  Jurkiewicz,
 Phys.Lett. { B278} (1992) 50. \\
M.E. Agishtein and A.A. Migdal, Mod. Phys. Lett. A7 (1992) 1039.\\
M.E. Agishtein and A.A. Migdal, Nucl.Phys. B385 (1992) 395.\\
J.Ambj\o rn,J.Jurkiewicz and C.F.Kristjansen. Nucl.Phys.
B393 (1993) 601.
\bibitem{many}J.Ambj\o rn,J.Jurkiewicz and C.F.Kristjansen. Nucl.Phys.
B393 (1993) 601.\\
B.Bruegmann and E.Marinari. Phys.Rev.Lett. 70 (1993) 1908.\\
B. Bruegmann, Phys. Rev. D47 (1993) 3330.\\
S.Catterall,J.Kogut and R.Renken. Phys.Lett. B328 (1994) 277.\\
B.V. DeBakker and J. Smit, Nucl.Phys. B439 (1995) 239.\\
J. Ambj\o rn and J. Jurkiewicz , Nucl.Phys. B451 (1996) 643.\\
P. Bialas, Z. Burda, A. Krzywicki and  B. Petersson, 
e-Print Archive: hep-lat/9601024 .


\bibitem{savvidy}G.K. Savvidy and  K.G. Savvidy,
Mod.Phys.Lett. A8 (1993) 2963;\\
Int. J. Mod. Phys. A8 (1993) 3993
\bibitem{ambarzum}R.V. Ambartzumian, G.K. Savvidy, K.G. Savvidy
and G.S. Sukiasian, Phys. Lett. B275 (1992) 99
\bibitem{savvidy1}G.K. Savvidy and K.G. Savvidy, Phys.Lett.
 B337 (1994) 333
\bibitem{savvidy2}G.K. Savvidy and K.G. Savvidy, Interaction hierarchy.
String and quantum gravity, Mod.Phys.Lett.A11 (1996) 1379, Hep-th 9506184
\bibitem{ass}J.Ambj\o rn,G.K. Savvidy and K.G. Savvidy. Alternative action
for quantum gravity and intrinsic rigidity of spacetime. NBI-HE-96-22
(hep-th 9606140) to appear in Nucl.Phys.B
\bibitem{durhuus}B.Durhuus and T.Jonsson.
Phys.Lett. B297 (1992) 271
\bibitem{johnston}C.F.Baillie and D.A.Johnston. Phys.Rev. D45 (1992) 3326
\bibitem{cappi} A.Cappi,P.Colangelo,G.Gonnella and A.Maritan.
Nucl.Phys.B370 (1992) 659


\bibitem{aw}C.B. Alendoerfer and A. Weil, Trans.Am.Soc. 55 (1943) 101-129.
\bibitem{herglotz}G.Herglotz, Abh. Math. Sem. Hansischen Univ. 15 (1943) 165
\bibitem{allendoerfer}C.B.Allendoerfer, Bull.Amer.Math.Soc. 54 (1948) 128
\bibitem{santalo}L.A.Santal\'o, Proc.Amer.Math.Soc. 1 (1950) 325
\bibitem{santalo1}L.A.Santal\'o, Revista Un.mat. Argentina 20 (1962) 79
\bibitem{schneider}R. Schneider, Geom.Dedicata 9 (1980) 111

\bibitem{poincare}H.Poincare, Calcul des probabilites,
2nd ed. Gauthier-Villars, Paris 1912
\bibitem{blaschke}W. Blaschke. "Integralgeometrie 17: Uber Kinematik", 
Deltion (Athens) 17 (1936) 3; "Integralgeometrie 17: Zur Kinematik",
Math. Z. 41 (1936) 465
\bibitem{chern}S.S.Chern and C.T.Yien. Bolletino della 
Unione  Matematica  Italiana  2 (1940) 434 ;
\bibitem{sch}G.K.Savvidy and R.Schneider. 
Comm. Math. Phys. 161 (1994) 283


\bibitem{steiner}J.Steiner, \"Uber parallele Fl\"achen, Gesammelte
Werke Band 2. (Berlin, 1882) S. 171-176
\bibitem{weyl}H.Weyl, Am.J.Math. 61 (1939) 461


\bibitem{che}J. Cheeger, W. M\"{u}lleer and R. Schrader, 
Comm.Math.Phys. 92 (1984) 405.

\bibitem{rw}M. Rocek and R.M. Williams, Phys.Lett. B273 (1991) 95.
\bibitem{hw1}H.W. Hamber and R.M. Williams, Nucl.Phys. B248 (1984) 392. 

\end{thebibliography}
\end{document}